\documentclass[amsmath,amssymb,aps,prb,twocolumn,floatfix,superscriptaddress]{revtex4-1}

\usepackage{graphicx}
\usepackage{dcolumn}
\usepackage{bm}
\usepackage{subfigure}
\usepackage{epstopdf}
\usepackage{enumerate}

\newcommand{\be}{\begin{equation}}
\newcommand{\ee}{\end{equation}}
\newcommand{\ba}{\begin{array}}
\newcommand{\ea}{\end{array}}
\newcommand{\bea}{\begin{eqnarray}} 
\newcommand{\eea}{\end{eqnarray}}

\def\temp(#1){\langle #1\rangle}

\def\tempp(#1){\langle {#1}|}

\def\temppp(#1){|#1\rangle}

\def\ttt(#1,#2){\left(\!\!\ba{c} {#1}\\{#2}\ea\!\!\right)}

\def\tttt(#1,#2){\left(\!\!\ba{cc} {#1} & {#2}\ea\!\!\right)}

\pdfpageattr {/Group << /S /Transparency /I true /CS /DeviceRGB>>}

\DeclareGraphicsExtensions{.pdf,.png}

\begin{document}

\title{Resonant single-parameter pumping in graphene}

\author{Y. Korniyenko}
\affiliation{Department of Microtechnology and Nanoscience - MC2,
Chalmers University of Technology, SE-412 96 G\"oteborg, Sweden}

\author{O. Shevtsov}
\affiliation{Department of Physics \& Astronomy, Northwestern University,
Evanston, Illinois 60208, USA}

\author{T. L\"ofwander}
\affiliation{Department of Microtechnology and Nanoscience - MC2,
Chalmers University of Technology, SE-412 96 G\"oteborg, Sweden}

\date{\today}

\begin{abstract}
We present results for non-adiabatic single-parameter pumping in a ballistic graphene field-effect transistor. We investigate how scattering from an ac-driven top gate results in dc charge current from source to drain in an asymmetric setup caused either by geometry of the device or different doping of leads. Charge current is computed using Floquet scattering matrix approach in Landauer-B\"uttiker operator formalism. We single out two mechanisms contributing to the pumped current: Fabry-P\'erot interference in open channels and quasibound state resonant scattering through closed channels. We identify two distinct parameter regimes based on the quasibound state scattering mechanism: high and low doping of contacts compared to the frequency of the ac drive. We show that the latter regime results in a stronger peak pump current. We discuss how back gate potential and temperature dependence can be used to change the direction of the pumped current, operating the device as a switch.
\end{abstract}

\maketitle

\section{Introduction}
Electrical pumping exploits asymmetry of a device under an ac field, generating a dc current between electrodes in the absence of voltage bias. The ac field breaks time-reversal symmetry, allowing for imbalance between charge carriers moving in opposite directions. In case the carriers preserve phase coherence during transport, the phenomenon is called quantum pumping and requires taking into account quantum interference in the device. A typical adiabatic quantum pump uses modulation of two ac voltages with a low frequency compared to the carrier traversal time across the device, but with a phase difference between them \cite{Moskalets02}, thus promoting charge transfer in one direction. It provides an interesting topological argument, relating closed cycles of the pump with generated electrical current \cite{Buttiker94,Switkes99,avron00}. A similar setup done with quantum dots, relying on Coulomb blockade, allows for quantized pumping, namely transfer of a single electron charge across the device per pump cycle. Quantized pumping is used to make single-electron sources with potential applications in metrology, namely development of a current standard \cite{Giblin16,Kaneko16}, and in electron quantum optics \cite{fletcher13,ubbelohde15}.

Since the discovery of graphene, it was brought to attention \cite{prada09} that decaying evanescent modes that normally do not contribute in pump setups for two-dimensional electron gases (2DEGs) play a big role around charge neutrality point of graphene. A unique operation mechanism based on promotion of these decaying states to propagating waves was proposed \cite{San-Jose11}, allowing for single-parameter pumping. Such setup requires maximal geometrical asymmetry of the device, with the ac gate being very close to one of the source/drain contacts. Nonadiabatic driving in this case results in a broadband evanescent mode promotion in contrast to resonance condition needed in corresponding 2DEG setup. Since then more setups were explored both for adiabatic \cite{low12} and non-adiabatic \cite{connolly13, san-jose12} pumping in graphene, also with potential applications in valleytronics \cite{jiang13,wang14}.

As the quality of ballistic graphene devices is improving \cite{Rickhaus2015,Chen2016,Zhao2015,Bandurin2016,Crossno2016,Ghahari2016}, resonance-assisted tunneling in graphene similar to setups in 2DEGs becomes possible. In the work presented here, we explore single-parameter bound-state-assisted non-adiabatic pumping in a ballistic transistor with the top gate positioned far away from both the source and the drain. In this setup, scattering via a bound state in the top gate barrier plays a major role in device operation. We demonstrate that asymmetry in the doping profile between the source and the drain and/or different distances between them and the top gate is enough to generate pumped current. We also show how the ambipolarity of the graphene band structure affects scattering through the bound state in graphene and how it can be used to tune the pumped current by changing the back gate voltage or changing the temperature of the system. We also explore how change in resonant scattering mechanism via the quasibound state between high and low doping of leads (compared with the pumping frequency) affects the pumped current.

\begin{figure}[b]
\includegraphics[width=\columnwidth]{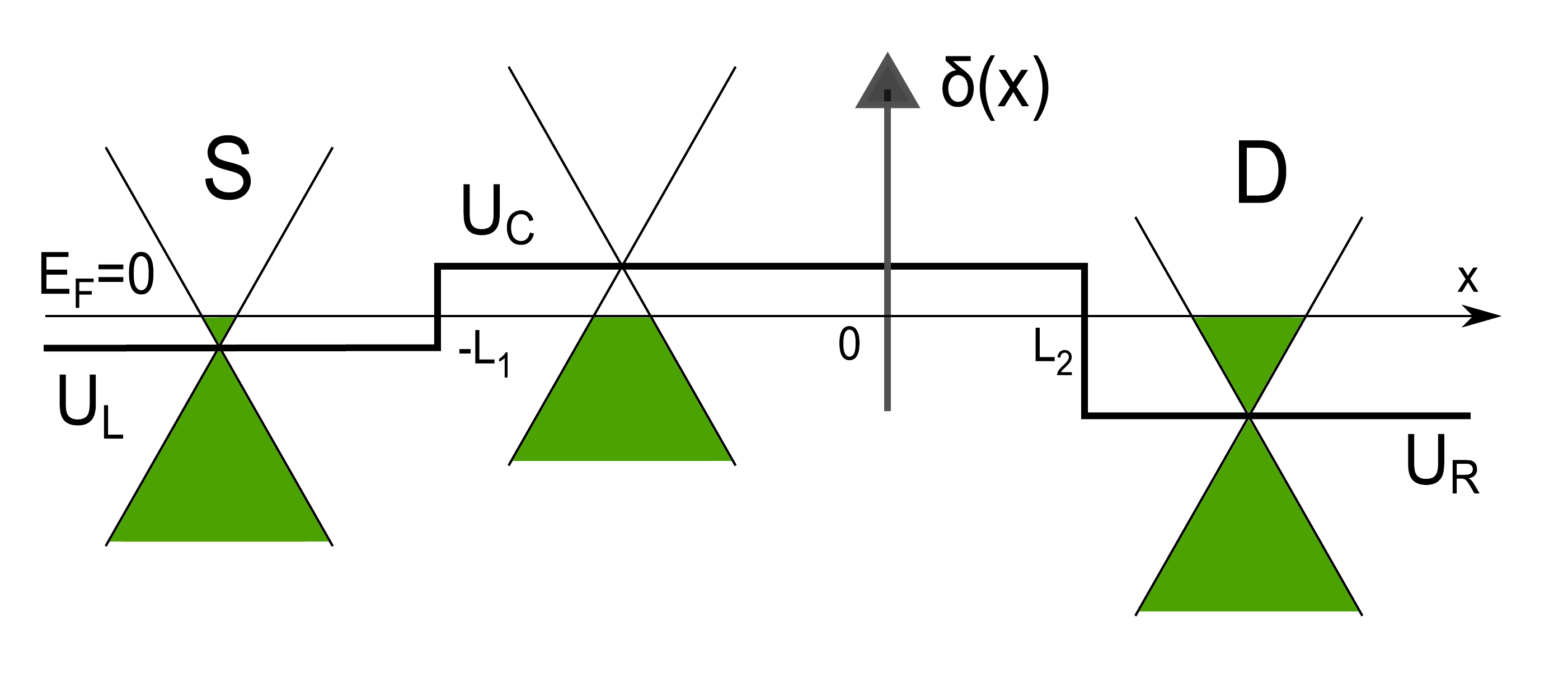}
\caption{Static potential landscape of the device, including asymmetric doping of the source (S) and drain (D) electrodes, and position of the top gate delta barrier. Dirac cones are at zero temperature filled up to the Fermi energy $E_F=0$. A back gate can tune the channel doping level $U_C$.}
\label{fig:device}
\end{figure}
%

\section{Model}\label{sec:model}

Our goal is to present a model of pumping in a transistor setup shown in Fig.~\ref{fig:device}. Source (S) and drain (D) electrodes deposited on top of graphene introduce finite doping, reflected respectively by potentials $U_L$ and $U_R$, which shift Dirac cones in the leads. The doping levels are therefore pinned under these contacts, and are assumed to be fixed in our theory. The potential $U_C$ in the central region $x\in [-L_1,L_2]$ can however be tuned by a back gate. The Fermi levels in the leads equilibrate and as a result there is a single Fermi energy across the device. We take it as the point of origin on the energy axis ($E_F=0$). Therefore, $U_C=0$ case corresponds to the Dirac point aligned with the Fermi energy, yielding the channel charge neutral in the absence of charge impurities.  The contacts are assumed to be ideal, so that the static potential landscape can be modeled using step functions
\begin{align}
U(x)=U_L\theta(-L_1-x)+U_R\theta(x-L_2) \nonumber\\
+U_C[\theta(x+L_1)-\theta(x-L_2)].
\end{align}
The potential steps are assumed to be smooth on the scale of the lattice spacing, but much sharper than the characteristic wavelength of the Dirac electrons $\lambda_D=\hbar v_f/|E-U_C|$, where $v_f\sim 10^6$~ m/s is the Fermi velocity. With the same argumentation, the narrow top gate potential can be modeled by a delta function, provided it is much wider than the graphene lattice spacing, while being much shorter than $\lambda_D$. Reformulating it in energy scales, we get $|E-U_C| \ll t$, where $t\sim2.8$ eV is the tight-binding hopping parameter. This requirement is satisfied within the linear Dirac cone section of the band structure. It allows us to also disregard any intervalley scattering and reduce the model Hamiltonian to only one valley, taking the form\cite{Korniyenko16, Korniyenko16_2, Korniyenko17}
\be
\mathcal{H}=-i\sigma_x\nabla_x+\sigma_y k_y+[Z_0+Z_1\cos (\Omega t)]\delta(x)+U(x).
\ee
Here we have set $\hbar=1$ and $v_F=1$. Pauli matrices $\sigma_{x,y}$ act in pseudospinor space. We assume the device to be wide enough in the transverse direction to disregard any edge effects or transverse quantization. For spatial homogeneity in the transverse direction, the transverse momentum $k_y$ is a good quantum number, and is conserved during scattering in the system. The top gate barrier is characterized by static ($Z_0$) and dynamic ($Z_1$) components with $\Omega$ being the frequency of the AC drive.

Periodicity of the Hamiltonian in time allows us to solve the time-dependent Dirac equation
\be
\mathcal{H}\psi(x,k_y,t)=i\partial_t\psi(x,k_y,t)
\ee
by using a Floquet ansatz, which is essentially a Fourier expansion in frequency harmonics
\be
\psi(x,k_y,t)=e^{-iEt}\sum_n \psi_n(x,k_y,E)e^{-in\Omega t}.
\ee
Effectively, time-periodicity of the drive generates wavefunction components $\psi_n$ at different sideband energies $E_n=E+n\Omega$. The ansatz allows us to rewrite the time-dependent Dirac equation as a matrix differential equation in energy sideband space and find solutions as shown in detail in our previous papers \cite{Korniyenko16, Korniyenko16_2, Korniyenko17}.
The solutions can be arranged into a Floquet scattering matrix $S_{\alpha\beta}(E_n,E_m)$, which describes scattering channels from electrode $\beta$ and energy $E_m$ to electrode $\alpha$ and energy $E_n$ [$\alpha,\beta\in\left\{L,R\right\}$ as in Fig.~\ref{fig:device}]. Landauer-B\"uttiker \cite{pedersen98,platero04,kohler05} operator approach is then used to express the pumped current in the absence of external source-drain bias. The induced pumped current in the drain is computed as [c.f. Eq.~(D10) in Ref.~\onlinecite{Korniyenko16}]
\be
I = \frac{2e}{\pi} \int\limits_{-\infty}^{\infty}  \sum_{k_y} \Delta T(E,k_y) f(E) dE,
\ee
%
where $f(E)=\left[1+\exp(E/\Xi)\right]^{-1}$ is the Fermi-Dirac distribution function, and $\Xi$ is the temperature (we set the Boltzmann constant $k_B=1$). The function
\be
\Delta T(E,k_y)=\sum_n\left[T_n(E,k_y)-T_n'(E,k_y)\right]
\ee
is the difference between transmission probabilities of scattering from source to drain and vice versa \footnote{$T_n$ are computed from Eqs.~(A16)-(A20) in Ref.~\cite{Korniyenko16_2}, while $T_n'$ are computed from a set of equations that are obtained from Eqs.~(A16)-(A20) by first changing sign of $\vec A$ and $\vec C$, thereafter exchanging $\bar\eta\leftrightarrow\eta$, and finally exchanging superscripts $L\leftrightarrow R$.}. Since the essential physics of scattering processes is contained in $\Delta T(E,k_y)$ we will proceed by first investigating this function as well as the energy-resolved current (after summing over $k_y$)
\be
I'(E)=\frac{2e}{\pi} \sum_{k_y} \Delta T(E,k_y),
\label{eq:current_kernel}
\ee
before presenting the temperature dependence of the total pumped current. In practice the sum over $k_y$ is turned into an integral over $k_y$ in the usual way. Since $\Delta T(E,k_y)$ is non-zero only for scattering between propagating waves in the leads, we find it practical to trade $k_y$ for an angle of incidence $\varphi$, through the relation
\be
k_y = |E-U_{\alpha}|\sin\varphi,
\ee
where for each energy we choose $\alpha=L$ or $R$ corresponding to the largest cone radius at this energy. In this way, the function $\Delta T(E,\varphi)$ is always defined for the whole interval $\varphi\in[-\pi/2,\pi/2]$.

\section{Results}\label{sec:results}
\subsection{Low doping}\label{sec:low}
%
%
%

\begin{figure}[t]
\includegraphics[width=0.9\columnwidth]{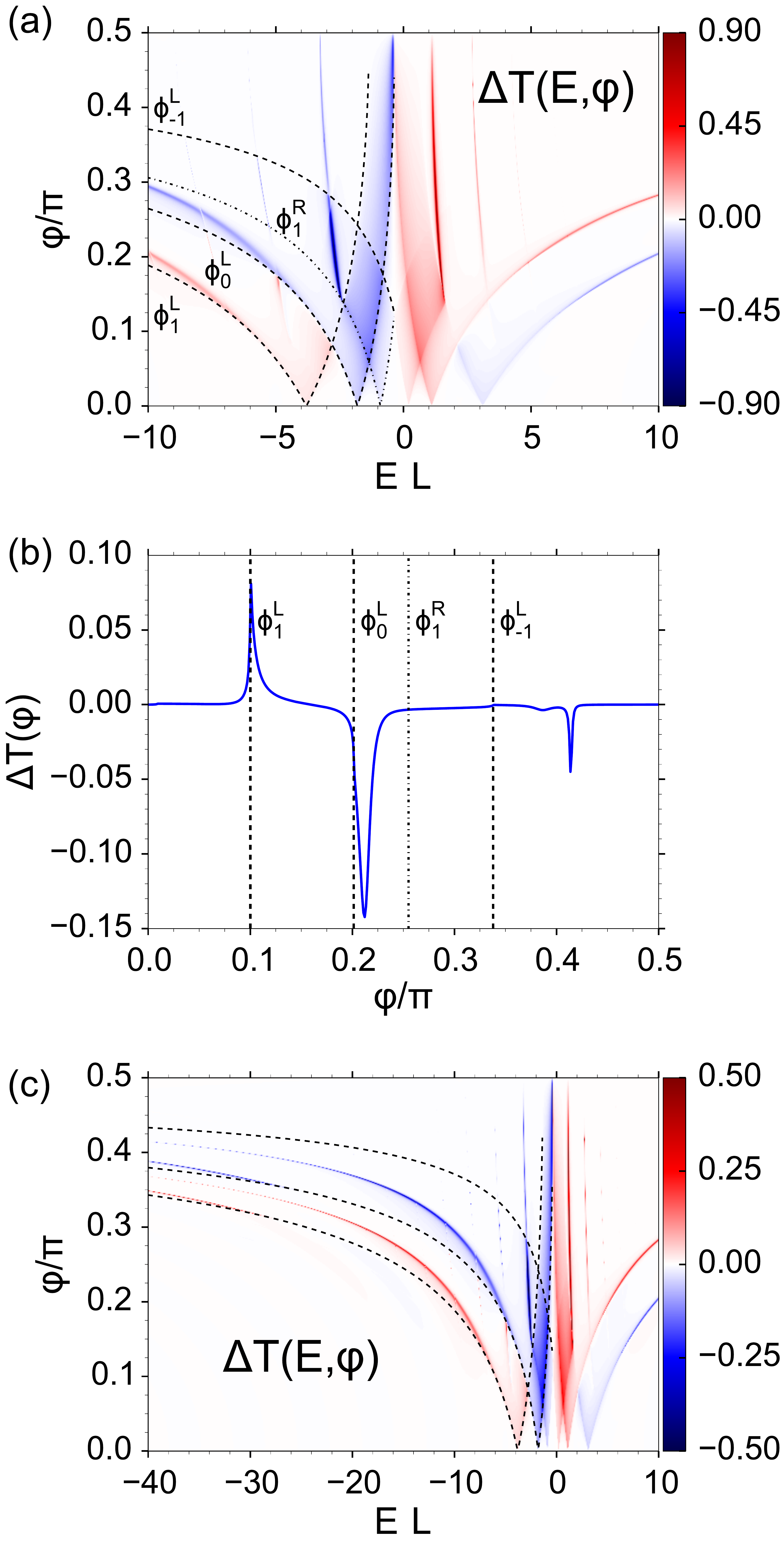}
\caption{Low doping regime ($U_L=-1.8L^{-1}$, $U_R=1.1L^{-1}$, $U_C=0$) with parameters $L_1=L_2=L/2$, $Z_0=0.4\pi$, $Z_1=0.45$, and $\Omega=2L^{-1}$. (a) Left-to-right total transmission probability imbalance. Dashed/dotted lines indicate evanescent wave region boundaries. (b) Cross section at E=-6/L in the map displayed in (a). (c) Fabry-P\'erot interference in open channels dictates transport at large $|E|$. }
\label{fig:2.1}
\end{figure}

\begin{figure*}[t]
\includegraphics[width=2\columnwidth]{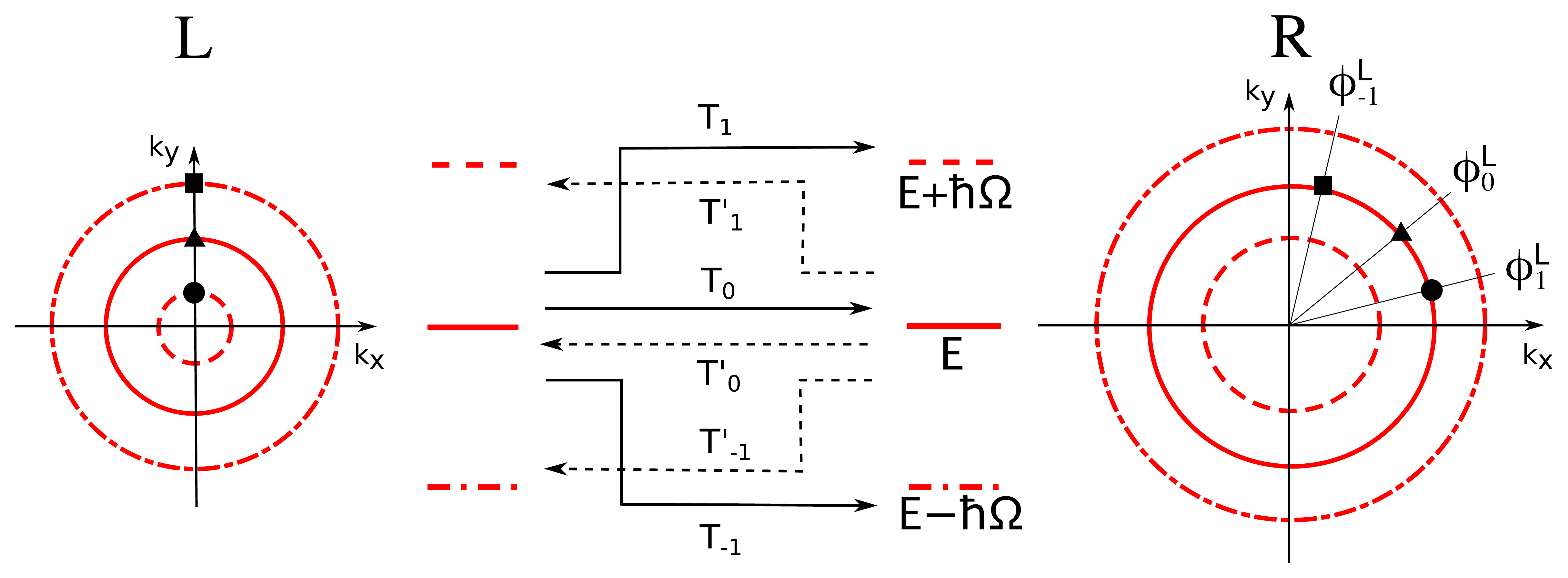}
\caption{Scattering processes responsible for abrupt switches in $\Delta T$ at critical angle boundaries in Fig.~\ref{fig:2.1}. Critical angles $\phi^L_n$ marked on the right lead Dirac cone cross section at energy E correspond to having no propagating states in the left lead at energies $E+n\Omega$ for $\varphi>\phi^L_n$ . The corresponding transmission processes are then forbidden.}
\label{fig:diagram}
\end{figure*}

\begin{figure*}[t]
\includegraphics[width=2\columnwidth]{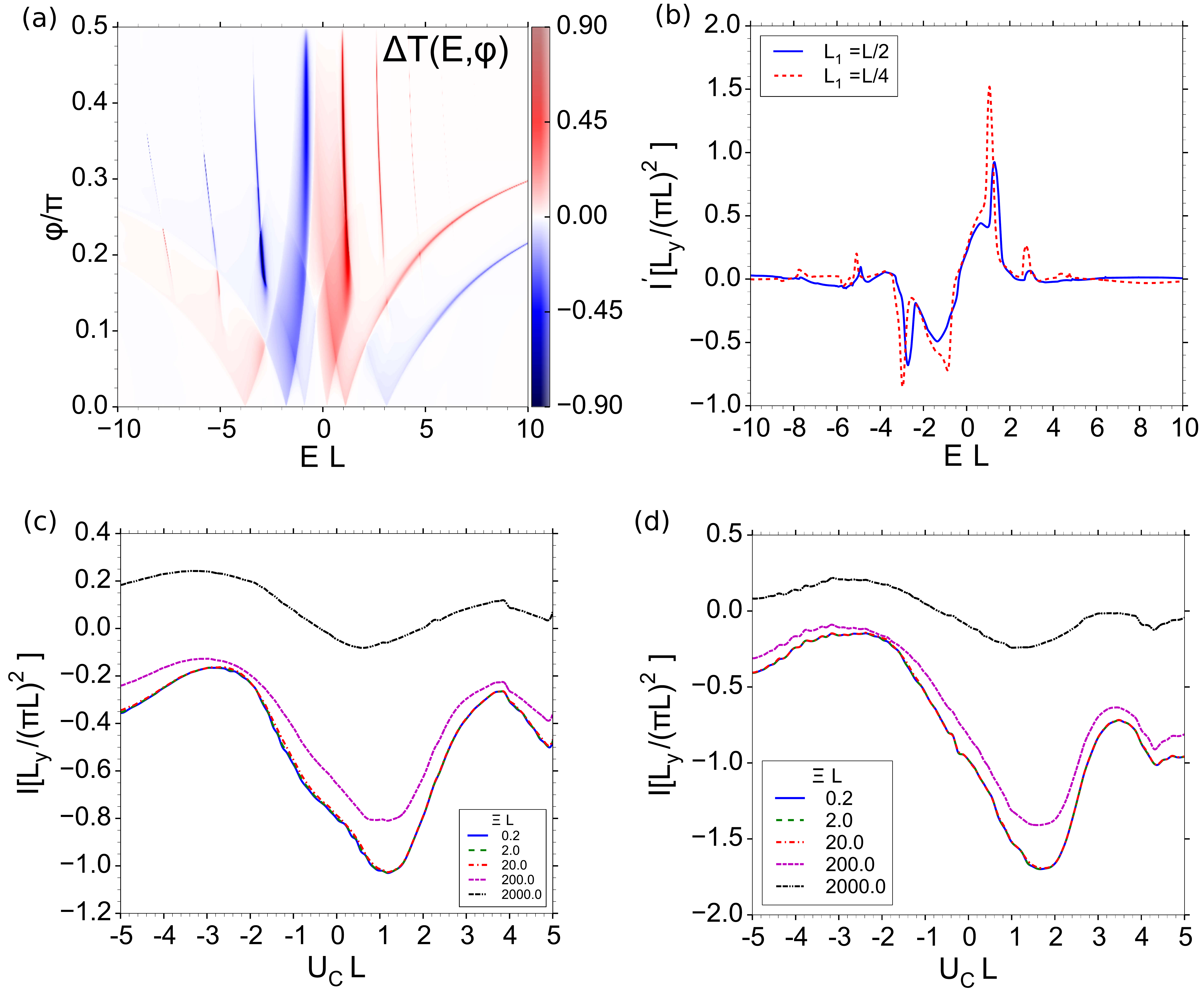}
%
%
%
%
\caption{(a) Left-to-right total transmission probability imbalance for parameters as in Fig.~\ref{fig:2.1}, but with an off-center delta barrier ($L_1=0.25L$). (b) Energy-resolved current for different top gate positions at $U_C=0$ and zero temperature. The total pump current at different temperatures for (c) the symmetric case $L_1=L_2$ , and for (d) the asymmetric case with $L_1=0.25L$.}
\label{fig:2.4}
\end{figure*}
%

The energy- and angle-resolved map of the transmission probability imbalance $\Delta T$ in the low doping regime $\Omega>U_{L,R}$ is shown in  Fig.~\ref{fig:2.1}(a). The figures display only positive angles, since the picture is symmetric with respect to $\varphi=0$. There is no reflection at angles $\varphi$ close to zero (perpendicular incidence), thus all transmission coefficients sum up to unity in both directions of propagation, effectively producing zero net current as is evident from the angle-resolved picture. The rest of the picture is characterized by two main features: 1) a background asymmetry $\Delta T(E,\varphi)$ with switches of the sign at boundaries between evanescent regions and 2) bound state resonances (seen as almost vertical lines) in those regions. The boundaries are described by a channel-dependent critical angle
\be
\phi^{\alpha\beta}_{n}(E)=\arcsin \left|\frac{E_n-U_{\alpha}}{E-U_\beta}\right|,
\ee
with waves being evanescent in that channel for any $|\varphi|>\phi^{\alpha\beta}_{n}(E)$, i.e. channel closed. Index $\beta$ indicates which of the L or R Dirac cones cross section is larger since angle $\varphi \in [-\pi/2,\pi/2]$ is defined in the larger cone. For the setup in Fig.~\ref{fig:2.1}, $\beta=R$ and we omit this index here on. It is instructive to look at the cross section of the map at a particular energy ($E=-6L^{-1}$) to understand the evolution of $\Delta T$ as angle of incidence changes, as shown in Fig.~\ref{fig:2.1}(b). Fig.~\ref{fig:diagram} displays the energy level diagram of the relevant processes. For small angles $|\varphi|<\phi_{1}^L$, all relevant transport channels are open and for small driving amplitude the difference between left-to-right and right-to-left transport is negligibly small. At $\varphi=\phi_{1}^L$ the wave vector in the left lead $k_1^L$ crosses into the imaginary domain, leaving this channel closed and $T'_1=0$. As the process $T_{-1}$ is still allowed, it results in a net probability current from left to right. For small driving, the corrections to $\Delta T$ are of order $Z_1^2$ in this case. The main band contribution is zero to the fourth order in the driving parameter, $T'_0=T_0+{O}(Z_1^4)$. As the angle is increased further to $\phi_{0}^L$, it leaves $k_0^L$ imaginary and forbids $T_{-1}$, $T_0$, $T_1$, and $T'_{0}$. There is only one relevant open channel left, $T'_{-1}$, resulting in a switch of direction of the probability current. For angles past $\phi_{-1}^{L}$ all those channels are closed and only contributions from second sidebands survive, thus only bound state resonances are visible. For energies closer to $U_R$ than $U_L$ (i.e. for $E>-0.35L^{-1}$ in Fig.~\ref{fig:2.1}), cone cross sections are larger in the left lead ($\beta=L$), and thus the relevant energy diagram is mirrored ($L \leftrightarrow R$) compared to Fig.~\ref{fig:diagram}, therefore $\Delta T$ switches sign as is seen in Fig.~\ref{fig:2.1}(a). Transmission via the bound state has been extensively studied in our previous works \cite{Korniyenko16, Korniyenko16_2}. The quasibound state contribution to $\Delta T$ comes from inelastic Breit-Wigner resonances in $T_{\pm n}$, $n>1$. Their strength is strictly decreasing with increasing $n$ in the weak-driving regime considered here, thus this contribution can be neglected for energies far away from the doping of the device. Indeed, as shown in Fig.~\ref{fig:2.1}(c), at energies $|E|\gg U_{L,R,C}$ only Fabry-P\'erot resonances and boundaries between open and closed channels contribute to the current. From the discussion above we conclude that the pumped current originates from an imbalance in inelastic scattering channels, notably the first sideband processes, and inelastic resonances in second sideband for $E$ close to $U_{L/R}$.

By shifting the position of the top gate one also moves the interference pattern in the device, but as long as evanescent wave amplitude reaching the delta barrier is negligible, the transmission map displays the same behavior as in the symmetric case, see Fig.~\ref{fig:2.4}(a), where $L_1=0.25L$. Angle integration of the current kernel [see Eq.~(\ref{eq:current_kernel})] reveals general dependence on energy as shown in Fig.~\ref{fig:2.4}(b). At energies far outside the range displayed in Fig.~\ref{fig:2.4}(b), the only surviving feature is the Fabry-P\'erot oscillations in open sideband channels, resulting in decaying oscillations in $I'(E)$. As any oscillating function, they cancel out to a large extent after integration over energies. The strongest contribution to the current comes from energies $U_L-N\Omega<E<U_R+N\Omega$ (N = number of sidebands), spanning the feature-rich region of panel Fig.~\ref{fig:2.4}(a). Shifting position of the gate results in change of the interference pattern, moving the resonances of interest as well, and thus also shifting strength between positive and negative contributions to current. $\Delta T$ as function of energy changes sign due to the asymmetry provided by the doping profile and it switches sign at $E=(U_L+U_R)/2$. However, the positive energy contribution is cut off by the Fermi function at low temperature, resulting in smaller positive contribution in the setup $U_L<U_R$. The total current after energy integration is therefore negative at low temperatures and charge neutral channel ($U_C=0$), see Fig.~\ref{fig:2.4}(c) and Fig.~\ref{fig:2.4}(d). In plots of $I'$ and $I$ we do not display constants $\hbar=v_f=e=1$ and only display scaling with the transverse width of the device $L_y$ and the normalization by the total length $L$ as a result of the energy variable given in units of $L^{-1}$. Reinstating the units, the largest current in Fig.~\ref{fig:2.4}(d) for device parameters $L=1 \, \mu$m and $L_y/L=10$ corresponds to 100 nA, which is of similar order of magnitude as results in the literature \cite{San-Jose11}. However, as the temperature is increased the peaks in $I'$ at positive energies are included and thus the total current also shifts towards positive values. In this sense, temperature acts as a parameter controlling the direction of pumped current flow.


Pumped current is zero in a symmetric setup when $U_L=U_R$ and $L_1=L_2$, while asymmetry in any of these parameter pairs results in a nonzero value. Comparing Fig.~\ref{fig:2.1}(a) and Fig.~\ref{fig:2.4}(a) we note that in the case of $L_1=0.25L_2$ the resonances are shifted to lower energies and negative peak contribution is stronger than in the symmetric setup. This is reflected in the total current magnitude, which is more negative as evident from Fig.~\ref{fig:2.4} (c) and (d). Thus if looking for a higher absolute value of pumped current, one has to increase the device asymmetry. On the other hand, increasing doping in the contacts brings the device into a high doping regime, where resonant scattering from a bound state happens via a different mechanism, as shown in the next section.

\subsection{High doping}\label{sec:high}
\begin{figure}[t]
\includegraphics[width=0.9\columnwidth]{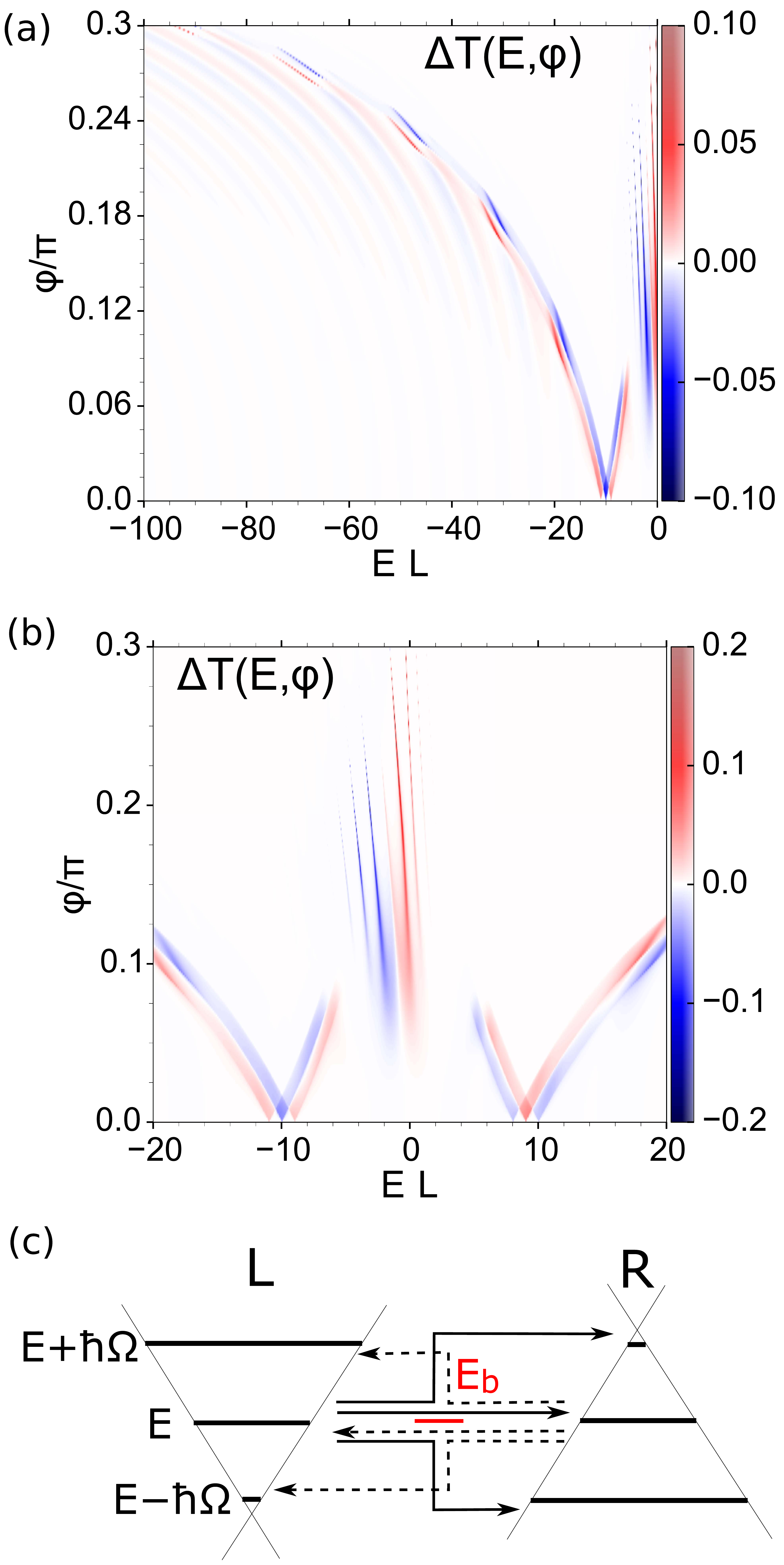}
\caption{High lead doping regime ($U_L=-10L^{-1}$, $U_R=9L^{-1}$, $U_C=0$) for a symmetric position of the top gate ($L_1=L_2=L/2$). The parameters are $Z_0=0.4\pi$, $Z_1=0.3$, and $\Omega=1L^{-1}$. (a) Color map including Fabry-P\'erot region, (b) color map in the relevant energy region, and (c) energy level diagram.}
\label{fig:1.1}
\end{figure}

Source-drain transmission imbalance for the case of high doping of contacts $|U_{L/R}|\gg|U_C|+\Omega$ and top gate in the symmetric position ($L_1=L_2=L/2$) is shown in Fig. \ref{fig:1.1}(a). For large positive or negative energies, similarly to the low-doping case, Fabry-P\'erot oscillations dominate the picture, gradually decreasing in amplitude. The key difference from the low doping scenario is contribution in the middle region $U_L<E<U_R$, see Fig.~\ref{fig:1.1}(b), where transport is mediated by evanescent waves. For a sufficiently long device $U_{L/R}\gg L^{-1}$, these evanescent waves do not reach from one lead to another thus contributing zero to overall transmission. The only possible way is to scatter via the bound state resulting in a process similar to tunneling through a double barrier, which has been explained in our previous work \cite{Korniyenko16_2}.
Consider the tunneling processes in the diagram in Fig.~\ref{fig:1.1}(c). The cone cross section sizes at energies $E_n$ are different in the left and right leads (due to the static doping profile). This results in transmission imbalances, for instance $\Delta T_1$ and $\Delta T_{-1}$ are non-zero, while the elastic contribution $\Delta T_0$ vanishes.
As a result, the resonant contribution to current is similar to that found in the low doping case, despite different formation mechanism of the resonances.

\begin{figure}[t]
\includegraphics[width=0.86\columnwidth]{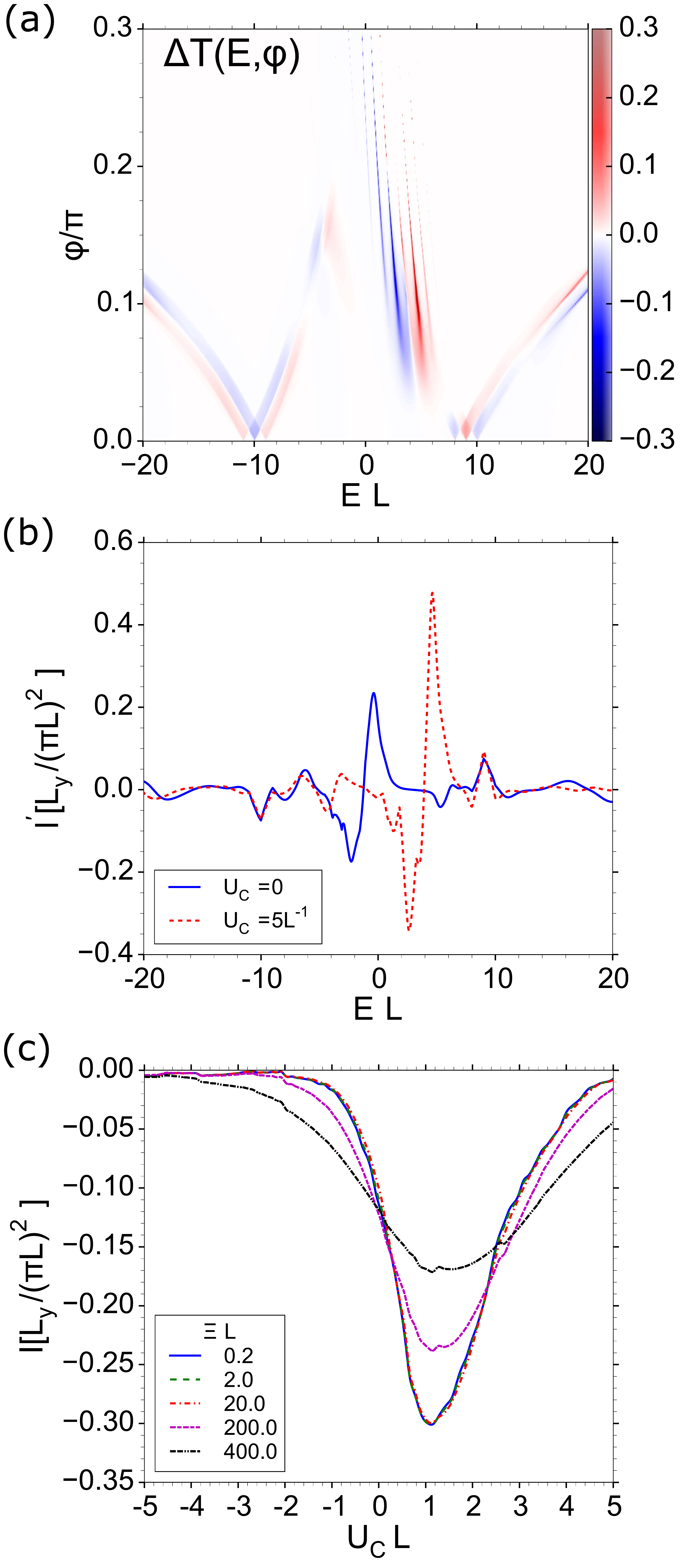}
\caption{High doping regime with parameters as in Fig.~\ref{fig:1.1} but with $U_C=5L^{-1}$. (a) Left-to-right total transmission probability imbalance. (b) Energy-resolved current at different channel dopings at zero temperature. (c) Total pump current as function of $U_C$ at different temperatures.}
\label{fig:1.2}
\end{figure}

Change in the channel doping shifts linearly the bound state position which in turn shifts the resonant peak contribution to $\Delta T$, see Fig.~\ref{fig:1.2}(a). As the pumped current strongly depends on the location of positive/negative peaks contribution with respect to the chemical potential, the back gate potential as well as temperature can be used to control the pumped current, as seen in Fig.~\ref{fig:1.2}(b) and Fig.~\ref{fig:1.2}(c), similarly to the low doping case. The difference between low and high doping regimes comes in the magnitude of the resonances. Since for the double-barrier tunneling the evanescent wave compensation is not exact for sidebands $n\neq0$ and decreases with sideband index, decaying due to difference in wave vectors as $\exp\left(-|k_n-k_0|L\right)$, while Breit-Wigner resonances in low doping case do not have this exponentially small prefactor, the latter tend to have higher amplitude, which also results in higher peak current as function of $U_C$.

\section{Summary}\label{sec:summary}
We have presented results describing a non-adiabatic pumping setup in a ballistic graphene field-effect transistor. We have shown how a static potential profile asymmetry together with time-reversal symmetry breaking caused by the ac gate potential result in source-drain pumped current. We have shown that the angle-resolved pumped current consists of Fabry-P\'erot oscillations with alternating sign due to opening/closing of scattering channels, with the effect decaying at low energies, and, on the other hand, quasibound state scattering resonances for energies close to the doping profile of the system [i.e. $E\sim U(x)$]. The pumped current for high channel transparencies is effectively the difference between the surface integrals of Dirac cones in the source and drain, thus asymmetry in them dictates the strength of the current. We have shown that for low doping of contacts with respect to the driving frequency, Fano- and Breit-Wigner type resonances result in a generally stronger peak pump current than double-barrier tunneling resonances for high contact doping. The high-doping regime requires the gate to be centered between the source and drain, while the low-doping pumped current is robust against gate shifts. We have shown that due to the ambipolarity of the dispersion relation of graphene, back gate potential and temperature can be used to change the direction of current, effectively turning the device into a switch.
\acknowledgments
We acknowledge financial support from the Swedish Foundation for Strategic Research, and the Knut and Alice Wallenberg Foundation. The research of O.S. was partly supported by the National Science Foundation (Grant No. DMR-1508730).


\begin{thebibliography}{28}%
\makeatletter
\providecommand \@ifxundefined [1]{%
 \@ifx{#1\undefined}
}%
\providecommand \@ifnum [1]{%
 \ifnum #1\expandafter \@firstoftwo
 \else \expandafter \@secondoftwo
 \fi
}%
\providecommand \@ifx [1]{%
 \ifx #1\expandafter \@firstoftwo
 \else \expandafter \@secondoftwo
 \fi
}%
\providecommand \natexlab [1]{#1}%
\providecommand \enquote  [1]{``#1''}%
\providecommand \bibnamefont  [1]{#1}%
\providecommand \bibfnamefont [1]{#1}%
\providecommand \citenamefont [1]{#1}%
\providecommand \href@noop [0]{\@secondoftwo}%
\providecommand \href [0]{\begingroup \@sanitize@url \@href}%
\providecommand \@href[1]{\@@startlink{#1}\@@href}%
\providecommand \@@href[1]{\endgroup#1\@@endlink}%
\providecommand \@sanitize@url [0]{\catcode `\\12\catcode `\$12\catcode
  `\&12\catcode `\#12\catcode `\^12\catcode `\_12\catcode `\%12\relax}%
\providecommand \@@startlink[1]{}%
\providecommand \@@endlink[0]{}%
\providecommand \url  [0]{\begingroup\@sanitize@url \@url }%
\providecommand \@url [1]{\endgroup\@href {#1}{\urlprefix }}%
\providecommand \urlprefix  [0]{URL }%
\providecommand \Eprint [0]{\href }%
\providecommand \doibase [0]{http://dx.doi.org/}%
\providecommand \selectlanguage [0]{\@gobble}%
\providecommand \bibinfo  [0]{\@secondoftwo}%
\providecommand \bibfield  [0]{\@secondoftwo}%
\providecommand \translation [1]{[#1]}%
\providecommand \BibitemOpen [0]{}%
\providecommand \bibitemStop [0]{}%
\providecommand \bibitemNoStop [0]{.\EOS\space}%
\providecommand \EOS [0]{\spacefactor3000\relax}%
\providecommand \BibitemShut  [1]{\csname bibitem#1\endcsname}%
\let\auto@bib@innerbib\@empty
\bibitem [{\citenamefont {Moskalets}\ and\ \citenamefont
  {B\"uttiker}(2002)}]{Moskalets02}%
  \BibitemOpen
  \bibfield  {author} {\bibinfo {author} {\bibfnamefont {M.}~\bibnamefont
  {Moskalets}}\ and\ \bibinfo {author} {\bibfnamefont {M.}~\bibnamefont
  {B\"uttiker}},\ }\bibfield  {title} {\enquote {\bibinfo {title} {Dissipation
  and noise in adiabatic quantum pumps},}\ }\href {\doibase
  10.1103/PhysRevB.66.035306} {\bibfield  {journal} {\bibinfo  {journal}
  {Physical Review B}\ }\textbf {\bibinfo {volume} {66}},\ \bibinfo {pages}
  {035306} (\bibinfo {year} {2002})}\BibitemShut {NoStop}%
\bibitem [{\citenamefont {B{\"u}ttiker}\ \emph {et~al.}(1994)\citenamefont
  {B{\"u}ttiker}, \citenamefont {Thomas},\ and\ \citenamefont
  {Pr{\^e}tre}}]{Buttiker94}%
  \BibitemOpen
  \bibfield  {author} {\bibinfo {author} {\bibfnamefont {M.}~\bibnamefont
  {B{\"u}ttiker}}, \bibinfo {author} {\bibfnamefont {H.}~\bibnamefont
  {Thomas}}, \ and\ \bibinfo {author} {\bibfnamefont {A.}~\bibnamefont
  {Pr{\^e}tre}},\ }\bibfield  {title} {\enquote {\bibinfo {title} {Current
  partition in multiprobe conductors in the presence of slowly oscillating
  external potentials},}\ }\href@noop {} {\bibfield  {journal} {\bibinfo
  {journal} {Zeitschrift f{\"u}r Physik B Condensed Matter}\ }\textbf {\bibinfo
  {volume} {94}},\ \bibinfo {pages} {133--137} (\bibinfo {year}
  {1994})}\BibitemShut {NoStop}%
\bibitem [{\citenamefont {Switkes}\ \emph {et~al.}(1999)\citenamefont
  {Switkes}, \citenamefont {Marcus}, \citenamefont {Campman},\ and\
  \citenamefont {Gossard}}]{Switkes99}%
  \BibitemOpen
  \bibfield  {author} {\bibinfo {author} {\bibfnamefont {M.}~\bibnamefont
  {Switkes}}, \bibinfo {author} {\bibfnamefont {C.~M.}\ \bibnamefont {Marcus}},
  \bibinfo {author} {\bibfnamefont {K.}~\bibnamefont {Campman}}, \ and\
  \bibinfo {author} {\bibfnamefont {A.~C.}\ \bibnamefont {Gossard}},\
  }\bibfield  {title} {\enquote {\bibinfo {title} {An adiabatic quantum
  electron pump},}\ }\href {\doibase 10.1126/science.283.5409.1905} {\bibfield
  {journal} {\bibinfo  {journal} {Science}\ }\textbf {\bibinfo {volume}
  {283}},\ \bibinfo {pages} {1905--1908} (\bibinfo {year} {1999})}\BibitemShut
  {NoStop}%
\bibitem [{\citenamefont {Avron}\ \emph {et~al.}(2000)\citenamefont {Avron},
  \citenamefont {Elgart}, \citenamefont {Graf},\ and\ \citenamefont
  {Sadun}}]{avron00}%
  \BibitemOpen
  \bibfield  {author} {\bibinfo {author} {\bibfnamefont {J.}~\bibnamefont
  {Avron}}, \bibinfo {author} {\bibfnamefont {A.}~\bibnamefont {Elgart}},
  \bibinfo {author} {\bibfnamefont {G.}~\bibnamefont {Graf}}, \ and\ \bibinfo
  {author} {\bibfnamefont {L.}~\bibnamefont {Sadun}},\ }\bibfield  {title}
  {\enquote {\bibinfo {title} {Geometry, statistics, and asymptotics of quantum
  pumps},}\ }\href@noop {} {\bibfield  {journal} {\bibinfo  {journal} {Physical
  Review B}\ }\textbf {\bibinfo {volume} {62}},\ \bibinfo {pages} {R10618}
  (\bibinfo {year} {2000})}\BibitemShut {NoStop}%
\bibitem [{\citenamefont {Giblin}\ \emph {et~al.}(2016)\citenamefont {Giblin},
  \citenamefont {See}, \citenamefont {Petrie}, \citenamefont {Janssen},
  \citenamefont {Farrer}, \citenamefont {Griffiths}, \citenamefont {Jones},
  \citenamefont {Ritchie},\ and\ \citenamefont {Kataoka}}]{Giblin16}%
  \BibitemOpen
  \bibfield  {author} {\bibinfo {author} {\bibfnamefont {S.}~\bibnamefont
  {Giblin}}, \bibinfo {author} {\bibfnamefont {P.}~\bibnamefont {See}},
  \bibinfo {author} {\bibfnamefont {A.}~\bibnamefont {Petrie}}, \bibinfo
  {author} {\bibfnamefont {T.}~\bibnamefont {Janssen}}, \bibinfo {author}
  {\bibfnamefont {I.}~\bibnamefont {Farrer}}, \bibinfo {author} {\bibfnamefont
  {J.}~\bibnamefont {Griffiths}}, \bibinfo {author} {\bibfnamefont
  {G.}~\bibnamefont {Jones}}, \bibinfo {author} {\bibfnamefont
  {D.}~\bibnamefont {Ritchie}}, \ and\ \bibinfo {author} {\bibfnamefont
  {M.}~\bibnamefont {Kataoka}},\ }\bibfield  {title} {\enquote {\bibinfo
  {title} {High-resolution error detection in the capture process of a
  single-electron pump},}\ }\href@noop {} {\bibfield  {journal} {\bibinfo
  {journal} {Applied Physics Letters}\ }\textbf {\bibinfo {volume} {108}},\
  \bibinfo {pages} {023502} (\bibinfo {year} {2016})}\BibitemShut {NoStop}%
\bibitem [{\citenamefont {Kaneko}\ \emph {et~al.}(2016)\citenamefont {Kaneko},
  \citenamefont {Nakamura},\ and\ \citenamefont {Okazaki}}]{Kaneko16}%
  \BibitemOpen
  \bibfield  {author} {\bibinfo {author} {\bibfnamefont {N.-H.}\ \bibnamefont
  {Kaneko}}, \bibinfo {author} {\bibfnamefont {S.}~\bibnamefont {Nakamura}}, \
  and\ \bibinfo {author} {\bibfnamefont {Y.}~\bibnamefont {Okazaki}},\
  }\bibfield  {title} {\enquote {\bibinfo {title} {A review of the quantum
  current standard},}\ }\href {http://stacks.iop.org/0957-0233/27/i=3/a=032001}
  {\bibfield  {journal} {\bibinfo  {journal} {Measurement Science and
  Technology}\ }\textbf {\bibinfo {volume} {27}},\ \bibinfo {pages} {032001}
  (\bibinfo {year} {2016})}\BibitemShut {NoStop}%
\bibitem [{\citenamefont {Fletcher}\ \emph {et~al.}(2013)\citenamefont
  {Fletcher}, \citenamefont {See}, \citenamefont {Howe}, \citenamefont
  {Pepper}, \citenamefont {Giblin}, \citenamefont {Griffiths}, \citenamefont
  {Jones}, \citenamefont {Farrer}, \citenamefont {Ritchie}, \citenamefont
  {Janssen} \emph {et~al.}}]{fletcher13}%
  \BibitemOpen
  \bibfield  {author} {\bibinfo {author} {\bibfnamefont {J.}~\bibnamefont
  {Fletcher}}, \bibinfo {author} {\bibfnamefont {P.}~\bibnamefont {See}},
  \bibinfo {author} {\bibfnamefont {H.}~\bibnamefont {Howe}}, \bibinfo {author}
  {\bibfnamefont {M.}~\bibnamefont {Pepper}}, \bibinfo {author} {\bibfnamefont
  {S.}~\bibnamefont {Giblin}}, \bibinfo {author} {\bibfnamefont
  {J.}~\bibnamefont {Griffiths}}, \bibinfo {author} {\bibfnamefont
  {G.}~\bibnamefont {Jones}}, \bibinfo {author} {\bibfnamefont
  {I.}~\bibnamefont {Farrer}}, \bibinfo {author} {\bibfnamefont
  {D.}~\bibnamefont {Ritchie}}, \bibinfo {author} {\bibfnamefont
  {T.}~\bibnamefont {Janssen}},  \emph {et~al.},\ }\bibfield  {title} {\enquote
  {\bibinfo {title} {Clock-controlled emission of single-electron wave packets
  in a solid-state circuit},}\ }\href@noop {} {\bibfield  {journal} {\bibinfo
  {journal} {Physical Review Letters}\ }\textbf {\bibinfo {volume} {111}},\
  \bibinfo {pages} {216807} (\bibinfo {year} {2013})}\BibitemShut {NoStop}%
\bibitem [{\citenamefont {Ubbelohde}\ \emph {et~al.}(2015)\citenamefont
  {Ubbelohde}, \citenamefont {Hohls}, \citenamefont {Kashcheyevs},
  \citenamefont {Wagner}, \citenamefont {Fricke}, \citenamefont {K{\"a}stner},
  \citenamefont {Pierz}, \citenamefont {Schumacher},\ and\ \citenamefont
  {Haug}}]{ubbelohde15}%
  \BibitemOpen
  \bibfield  {author} {\bibinfo {author} {\bibfnamefont {N.}~\bibnamefont
  {Ubbelohde}}, \bibinfo {author} {\bibfnamefont {F.}~\bibnamefont {Hohls}},
  \bibinfo {author} {\bibfnamefont {V.}~\bibnamefont {Kashcheyevs}}, \bibinfo
  {author} {\bibfnamefont {T.}~\bibnamefont {Wagner}}, \bibinfo {author}
  {\bibfnamefont {L.}~\bibnamefont {Fricke}}, \bibinfo {author} {\bibfnamefont
  {B.}~\bibnamefont {K{\"a}stner}}, \bibinfo {author} {\bibfnamefont
  {K.}~\bibnamefont {Pierz}}, \bibinfo {author} {\bibfnamefont {H.~W.}\
  \bibnamefont {Schumacher}}, \ and\ \bibinfo {author} {\bibfnamefont {R.~J.}\
  \bibnamefont {Haug}},\ }\bibfield  {title} {\enquote {\bibinfo {title}
  {Partitioning of on-demand electron pairs},}\ }\href@noop {} {\bibfield
  {journal} {\bibinfo  {journal} {Nature Nanotechnology}\ }\textbf {\bibinfo
  {volume} {10}},\ \bibinfo {pages} {46--49} (\bibinfo {year}
  {2015})}\BibitemShut {NoStop}%
\bibitem [{\citenamefont {Prada}\ \emph {et~al.}(2009)\citenamefont {Prada},
  \citenamefont {San-Jose},\ and\ \citenamefont {Schomerus}}]{prada09}%
  \BibitemOpen
  \bibfield  {author} {\bibinfo {author} {\bibfnamefont {E.}~\bibnamefont
  {Prada}}, \bibinfo {author} {\bibfnamefont {P.}~\bibnamefont {San-Jose}}, \
  and\ \bibinfo {author} {\bibfnamefont {H.}~\bibnamefont {Schomerus}},\
  }\bibfield  {title} {\enquote {\bibinfo {title} {Quantum pumping in
  graphene},}\ }\href@noop {} {\bibfield  {journal} {\bibinfo  {journal}
  {Physical Review B}\ }\textbf {\bibinfo {volume} {80}},\ \bibinfo {pages}
  {245414} (\bibinfo {year} {2009})}\BibitemShut {NoStop}%
\bibitem [{\citenamefont {San-Jose}\ \emph {et~al.}(2011)\citenamefont
  {San-Jose}, \citenamefont {Prada}, \citenamefont {Kohler},\ and\
  \citenamefont {Schomerus}}]{San-Jose11}%
  \BibitemOpen
  \bibfield  {author} {\bibinfo {author} {\bibfnamefont {P.}~\bibnamefont
  {San-Jose}}, \bibinfo {author} {\bibfnamefont {E.}~\bibnamefont {Prada}},
  \bibinfo {author} {\bibfnamefont {S.}~\bibnamefont {Kohler}}, \ and\ \bibinfo
  {author} {\bibfnamefont {H.}~\bibnamefont {Schomerus}},\ }\bibfield  {title}
  {\enquote {\bibinfo {title} {Single-parameter pumping in graphene},}\ }\href
  {\doibase 10.1103/PhysRevB.84.155408} {\bibfield  {journal} {\bibinfo
  {journal} {Physical Review B}\ }\textbf {\bibinfo {volume} {84}},\ \bibinfo
  {pages} {155408} (\bibinfo {year} {2011})}\BibitemShut {NoStop}%
\bibitem [{\citenamefont {Low}\ \emph {et~al.}(2012)\citenamefont {Low},
  \citenamefont {Jiang}, \citenamefont {Katsnelson},\ and\ \citenamefont
  {Guinea}}]{low12}%
  \BibitemOpen
  \bibfield  {author} {\bibinfo {author} {\bibfnamefont {T.}~\bibnamefont
  {Low}}, \bibinfo {author} {\bibfnamefont {Y.}~\bibnamefont {Jiang}}, \bibinfo
  {author} {\bibfnamefont {M.}~\bibnamefont {Katsnelson}}, \ and\ \bibinfo
  {author} {\bibfnamefont {F.}~\bibnamefont {Guinea}},\ }\bibfield  {title}
  {\enquote {\bibinfo {title} {Electron pumping in graphene mechanical
  resonators},}\ }\href@noop {} {\bibfield  {journal} {\bibinfo  {journal}
  {Nano Letters}\ }\textbf {\bibinfo {volume} {12}},\ \bibinfo {pages}
  {850--854} (\bibinfo {year} {2012})}\BibitemShut {NoStop}%
\bibitem [{\citenamefont {Connolly}\ \emph {et~al.}(2013)\citenamefont
  {Connolly}, \citenamefont {Chiu}, \citenamefont {Giblin}, \citenamefont
  {Kataoka}, \citenamefont {Fletcher}, \citenamefont {Chua}, \citenamefont
  {Griffiths}, \citenamefont {Jones}, \citenamefont {Fal'Ko}, \citenamefont
  {Smith} \emph {et~al.}}]{connolly13}%
  \BibitemOpen
  \bibfield  {author} {\bibinfo {author} {\bibfnamefont {M.}~\bibnamefont
  {Connolly}}, \bibinfo {author} {\bibfnamefont {K.}~\bibnamefont {Chiu}},
  \bibinfo {author} {\bibfnamefont {S.}~\bibnamefont {Giblin}}, \bibinfo
  {author} {\bibfnamefont {M.}~\bibnamefont {Kataoka}}, \bibinfo {author}
  {\bibfnamefont {J.}~\bibnamefont {Fletcher}}, \bibinfo {author}
  {\bibfnamefont {C.}~\bibnamefont {Chua}}, \bibinfo {author} {\bibfnamefont
  {J.}~\bibnamefont {Griffiths}}, \bibinfo {author} {\bibfnamefont
  {G.}~\bibnamefont {Jones}}, \bibinfo {author} {\bibfnamefont
  {V.}~\bibnamefont {Fal'Ko}}, \bibinfo {author} {\bibfnamefont
  {C.}~\bibnamefont {Smith}},  \emph {et~al.},\ }\bibfield  {title} {\enquote
  {\bibinfo {title} {Gigahertz quantized charge pumping in graphene quantum
  dots},}\ }\href@noop {} {\bibfield  {journal} {\bibinfo  {journal} {Nature
  Nanotechnology}\ }\textbf {\bibinfo {volume} {8}},\ \bibinfo {pages}
  {417--420} (\bibinfo {year} {2013})}\BibitemShut {NoStop}%
\bibitem [{\citenamefont {San-Jose}\ \emph {et~al.}(2012)\citenamefont
  {San-Jose}, \citenamefont {Prada}, \citenamefont {Schomerus},\ and\
  \citenamefont {Kohler}}]{san-jose12}%
  \BibitemOpen
  \bibfield  {author} {\bibinfo {author} {\bibfnamefont {P.}~\bibnamefont
  {San-Jose}}, \bibinfo {author} {\bibfnamefont {E.}~\bibnamefont {Prada}},
  \bibinfo {author} {\bibfnamefont {H.}~\bibnamefont {Schomerus}}, \ and\
  \bibinfo {author} {\bibfnamefont {S.}~\bibnamefont {Kohler}},\ }\bibfield
  {title} {\enquote {\bibinfo {title} {Laser-induced quantum pumping in
  graphene},}\ }\href@noop {} {\bibfield  {journal} {\bibinfo  {journal}
  {Applied Physics Letters}\ }\textbf {\bibinfo {volume} {101}},\ \bibinfo
  {pages} {153506} (\bibinfo {year} {2012})}\BibitemShut {NoStop}%
\bibitem [{\citenamefont {Jiang}\ \emph {et~al.}(2013)\citenamefont {Jiang},
  \citenamefont {Low}, \citenamefont {Chang}, \citenamefont {Katsnelson},\ and\
  \citenamefont {Guinea}}]{jiang13}%
  \BibitemOpen
  \bibfield  {author} {\bibinfo {author} {\bibfnamefont {Y.}~\bibnamefont
  {Jiang}}, \bibinfo {author} {\bibfnamefont {T.}~\bibnamefont {Low}}, \bibinfo
  {author} {\bibfnamefont {K.}~\bibnamefont {Chang}}, \bibinfo {author}
  {\bibfnamefont {M.~I.}\ \bibnamefont {Katsnelson}}, \ and\ \bibinfo {author}
  {\bibfnamefont {F.}~\bibnamefont {Guinea}},\ }\bibfield  {title} {\enquote
  {\bibinfo {title} {Generation of pure bulk valley current in graphene},}\
  }\href@noop {} {\bibfield  {journal} {\bibinfo  {journal} {Physical Review
  Letters}\ }\textbf {\bibinfo {volume} {110}},\ \bibinfo {pages} {046601}
  (\bibinfo {year} {2013})}\BibitemShut {NoStop}%
\bibitem [{\citenamefont {Wang}\ \emph {et~al.}(2014)\citenamefont {Wang},
  \citenamefont {Chan},\ and\ \citenamefont {Lin}}]{wang14}%
  \BibitemOpen
  \bibfield  {author} {\bibinfo {author} {\bibfnamefont {J.}~\bibnamefont
  {Wang}}, \bibinfo {author} {\bibfnamefont {K.}~\bibnamefont {Chan}}, \ and\
  \bibinfo {author} {\bibfnamefont {Z.}~\bibnamefont {Lin}},\ }\bibfield
  {title} {\enquote {\bibinfo {title} {Quantum pumping of valley current in
  strain engineered graphene},}\ }\href@noop {} {\bibfield  {journal} {\bibinfo
   {journal} {Applied Physics Letters}\ }\textbf {\bibinfo {volume} {104}},\
  \bibinfo {pages} {013105} (\bibinfo {year} {2014})}\BibitemShut {NoStop}%
\bibitem [{\citenamefont {Rickhaus}\ \emph {et~al.}(2015)\citenamefont
  {Rickhaus}, \citenamefont {Makk}, \citenamefont {Liu}, \citenamefont
  {T{\'o}v{\'a}ri}, \citenamefont {Weiss}, \citenamefont {Maurand},
  \citenamefont {Richter},\ and\ \citenamefont
  {Schoenenberger}}]{Rickhaus2015}%
  \BibitemOpen
  \bibfield  {author} {\bibinfo {author} {\bibfnamefont {P.}~\bibnamefont
  {Rickhaus}}, \bibinfo {author} {\bibfnamefont {P.}~\bibnamefont {Makk}},
  \bibinfo {author} {\bibfnamefont {M.-H.}\ \bibnamefont {Liu}}, \bibinfo
  {author} {\bibfnamefont {E.}~\bibnamefont {T{\'o}v{\'a}ri}}, \bibinfo
  {author} {\bibfnamefont {M.}~\bibnamefont {Weiss}}, \bibinfo {author}
  {\bibfnamefont {R.}~\bibnamefont {Maurand}}, \bibinfo {author} {\bibfnamefont
  {K.}~\bibnamefont {Richter}}, \ and\ \bibinfo {author} {\bibfnamefont
  {C.}~\bibnamefont {Schoenenberger}},\ }\bibfield  {title} {\enquote {\bibinfo
  {title} {{Snake trajectories in ultraclean graphene p-n junctions}},}\
  }\href@noop {} {\bibfield  {journal} {\bibinfo  {journal} {Nature
  Communications}\ }\textbf {\bibinfo {volume} {6}} (\bibinfo {year}
  {2015})}\BibitemShut {NoStop}%
\bibitem [{\citenamefont {Chen}\ \emph {et~al.}(2016)\citenamefont {Chen},
  \citenamefont {Han}, \citenamefont {Elahi}, \citenamefont {Habib},
  \citenamefont {Wang}, \citenamefont {Wen}, \citenamefont {Gao}, \citenamefont
  {Taniguchi}, \citenamefont {Watanabe}, \citenamefont {Hone}, \citenamefont
  {Ghosh},\ and\ \citenamefont {Dean}}]{Chen2016}%
  \BibitemOpen
  \bibfield  {author} {\bibinfo {author} {\bibfnamefont {S.}~\bibnamefont
  {Chen}}, \bibinfo {author} {\bibfnamefont {Z.}~\bibnamefont {Han}}, \bibinfo
  {author} {\bibfnamefont {M.~M.}\ \bibnamefont {Elahi}}, \bibinfo {author}
  {\bibfnamefont {K.~M.~M.}\ \bibnamefont {Habib}}, \bibinfo {author}
  {\bibfnamefont {L.}~\bibnamefont {Wang}}, \bibinfo {author} {\bibfnamefont
  {B.}~\bibnamefont {Wen}}, \bibinfo {author} {\bibfnamefont {Y.}~\bibnamefont
  {Gao}}, \bibinfo {author} {\bibfnamefont {T.}~\bibnamefont {Taniguchi}},
  \bibinfo {author} {\bibfnamefont {K.}~\bibnamefont {Watanabe}}, \bibinfo
  {author} {\bibfnamefont {J.}~\bibnamefont {Hone}}, \bibinfo {author}
  {\bibfnamefont {A.~W.}\ \bibnamefont {Ghosh}}, \ and\ \bibinfo {author}
  {\bibfnamefont {C.~R.}\ \bibnamefont {Dean}},\ }\bibfield  {title} {\enquote
  {\bibinfo {title} {{Electron optics with p-n junctions in ballistic
  graphene.}}}\ }\href@noop {} {\bibfield  {journal} {\bibinfo  {journal}
  {Science (New York, NY)}\ }\textbf {\bibinfo {volume} {353}},\ \bibinfo
  {pages} {1522--1525} (\bibinfo {year} {2016})}\BibitemShut {NoStop}%
\bibitem [{\citenamefont {Zhao}\ \emph {et~al.}(2015)\citenamefont {Zhao},
  \citenamefont {Wyrick}, \citenamefont {Natterer}, \citenamefont
  {Rodriguez-Nieva}, \citenamefont {Lewandowski}, \citenamefont {Watanabe},
  \citenamefont {Taniguchi}, \citenamefont {Levitov}, \citenamefont
  {Zhitenev},\ and\ \citenamefont {Stroscio}}]{Zhao2015}%
  \BibitemOpen
  \bibfield  {author} {\bibinfo {author} {\bibfnamefont {Y.}~\bibnamefont
  {Zhao}}, \bibinfo {author} {\bibfnamefont {J.}~\bibnamefont {Wyrick}},
  \bibinfo {author} {\bibfnamefont {F.~D.}\ \bibnamefont {Natterer}}, \bibinfo
  {author} {\bibfnamefont {J.~F.}\ \bibnamefont {Rodriguez-Nieva}}, \bibinfo
  {author} {\bibfnamefont {C.}~\bibnamefont {Lewandowski}}, \bibinfo {author}
  {\bibfnamefont {K.}~\bibnamefont {Watanabe}}, \bibinfo {author}
  {\bibfnamefont {T.}~\bibnamefont {Taniguchi}}, \bibinfo {author}
  {\bibfnamefont {L.~S.}\ \bibnamefont {Levitov}}, \bibinfo {author}
  {\bibfnamefont {N.~B.}\ \bibnamefont {Zhitenev}}, \ and\ \bibinfo {author}
  {\bibfnamefont {J.~A.}\ \bibnamefont {Stroscio}},\ }\bibfield  {title}
  {\enquote {\bibinfo {title} {{Creating and probing electron
  whispering-gallery modes in graphene}},}\ }\href@noop {} {\bibfield
  {journal} {\bibinfo  {journal} {Science (New York, NY)}\ }\textbf {\bibinfo
  {volume} {348}},\ \bibinfo {pages} {672--675} (\bibinfo {year}
  {2015})}\BibitemShut {NoStop}%
\bibitem [{\citenamefont {Bandurin}\ \emph {et~al.}(2016)\citenamefont
  {Bandurin}, \citenamefont {Torre}, \citenamefont {Kumar}, \citenamefont
  {Ben~Shalom}, \citenamefont {Tomadin}, \citenamefont {Principi},
  \citenamefont {Auton}, \citenamefont {Khestanova}, \citenamefont {Novoselov},
  \citenamefont {Grigorieva}, \citenamefont {Ponomarenko}, \citenamefont
  {Geim},\ and\ \citenamefont {Polini}}]{Bandurin2016}%
  \BibitemOpen
  \bibfield  {author} {\bibinfo {author} {\bibfnamefont {D.~A.}\ \bibnamefont
  {Bandurin}}, \bibinfo {author} {\bibfnamefont {I.}~\bibnamefont {Torre}},
  \bibinfo {author} {\bibfnamefont {R.~K.}\ \bibnamefont {Kumar}}, \bibinfo
  {author} {\bibfnamefont {M.}~\bibnamefont {Ben~Shalom}}, \bibinfo {author}
  {\bibfnamefont {A.}~\bibnamefont {Tomadin}}, \bibinfo {author} {\bibfnamefont
  {A.}~\bibnamefont {Principi}}, \bibinfo {author} {\bibfnamefont {G.~H.}\
  \bibnamefont {Auton}}, \bibinfo {author} {\bibfnamefont {E.}~\bibnamefont
  {Khestanova}}, \bibinfo {author} {\bibfnamefont {K.~S.}\ \bibnamefont
  {Novoselov}}, \bibinfo {author} {\bibfnamefont {I.~V.}\ \bibnamefont
  {Grigorieva}}, \bibinfo {author} {\bibfnamefont {L.~A.}\ \bibnamefont
  {Ponomarenko}}, \bibinfo {author} {\bibfnamefont {A.~K.}\ \bibnamefont
  {Geim}}, \ and\ \bibinfo {author} {\bibfnamefont {M.}~\bibnamefont
  {Polini}},\ }\bibfield  {title} {\enquote {\bibinfo {title} {{Negative local
  resistance caused by viscous electron backflow in graphene}},}\ }\href@noop
  {} {\bibfield  {journal} {\bibinfo  {journal} {Science (New York, NY)}\
  }\textbf {\bibinfo {volume} {351}},\ \bibinfo {pages} {1055--1058} (\bibinfo
  {year} {2016})}\BibitemShut {NoStop}%
\bibitem [{\citenamefont {Crossno}\ \emph {et~al.}(2016)\citenamefont
  {Crossno}, \citenamefont {Shi}, \citenamefont {Wang}, \citenamefont {Liu},
  \citenamefont {Harzheim}, \citenamefont {Lucas}, \citenamefont {Sachdev},
  \citenamefont {Kim}, \citenamefont {Taniguchi}, \citenamefont {Watanabe},
  \citenamefont {Ohki},\ and\ \citenamefont {Fong}}]{Crossno2016}%
  \BibitemOpen
  \bibfield  {author} {\bibinfo {author} {\bibfnamefont {J.}~\bibnamefont
  {Crossno}}, \bibinfo {author} {\bibfnamefont {J.~K.}\ \bibnamefont {Shi}},
  \bibinfo {author} {\bibfnamefont {K.}~\bibnamefont {Wang}}, \bibinfo {author}
  {\bibfnamefont {X.}~\bibnamefont {Liu}}, \bibinfo {author} {\bibfnamefont
  {A.}~\bibnamefont {Harzheim}}, \bibinfo {author} {\bibfnamefont
  {A.}~\bibnamefont {Lucas}}, \bibinfo {author} {\bibfnamefont
  {S.}~\bibnamefont {Sachdev}}, \bibinfo {author} {\bibfnamefont
  {P.}~\bibnamefont {Kim}}, \bibinfo {author} {\bibfnamefont {T.}~\bibnamefont
  {Taniguchi}}, \bibinfo {author} {\bibfnamefont {K.}~\bibnamefont {Watanabe}},
  \bibinfo {author} {\bibfnamefont {T.~A.}\ \bibnamefont {Ohki}}, \ and\
  \bibinfo {author} {\bibfnamefont {K.~C.}\ \bibnamefont {Fong}},\ }\bibfield
  {title} {\enquote {\bibinfo {title} {{Observation of the Dirac fluid and the
  breakdown of the Wiedemann-Franz law in graphene}},}\ }\href@noop {}
  {\bibfield  {journal} {\bibinfo  {journal} {Science (New York, NY)}\ }\textbf
  {\bibinfo {volume} {351}},\ \bibinfo {pages} {1058--1061} (\bibinfo {year}
  {2016})}\BibitemShut {NoStop}%
\bibitem [{\citenamefont {Ghahari}\ \emph {et~al.}(2016)\citenamefont
  {Ghahari}, \citenamefont {Xie}, \citenamefont {Taniguchi}, \citenamefont
  {Watanabe}, \citenamefont {Foster},\ and\ \citenamefont {Kim}}]{Ghahari2016}%
  \BibitemOpen
  \bibfield  {author} {\bibinfo {author} {\bibfnamefont {F.}~\bibnamefont
  {Ghahari}}, \bibinfo {author} {\bibfnamefont {H.-Y.}\ \bibnamefont {Xie}},
  \bibinfo {author} {\bibfnamefont {T.}~\bibnamefont {Taniguchi}}, \bibinfo
  {author} {\bibfnamefont {K.}~\bibnamefont {Watanabe}}, \bibinfo {author}
  {\bibfnamefont {M.~S.}\ \bibnamefont {Foster}}, \ and\ \bibinfo {author}
  {\bibfnamefont {P.}~\bibnamefont {Kim}},\ }\bibfield  {title} {\enquote
  {\bibinfo {title} {{Enhanced Thermoelectric Power in Graphene: Violation of
  the Mott Relation by Inelastic Scattering}},}\ }\href@noop {} {\bibfield
  {journal} {\bibinfo  {journal} {Physical Review Letters}\ }\textbf {\bibinfo
  {volume} {116}},\ \bibinfo {pages} {136802} (\bibinfo {year}
  {2016})}\BibitemShut {NoStop}%
\bibitem [{\citenamefont {Korniyenko}\ \emph
  {et~al.}(2016{\natexlab{a}})\citenamefont {Korniyenko}, \citenamefont
  {Shevtsov},\ and\ \citenamefont {L\"ofwander}}]{Korniyenko16}%
  \BibitemOpen
  \bibfield  {author} {\bibinfo {author} {\bibfnamefont {Y.}~\bibnamefont
  {Korniyenko}}, \bibinfo {author} {\bibfnamefont {O.}~\bibnamefont
  {Shevtsov}}, \ and\ \bibinfo {author} {\bibfnamefont {T.}~\bibnamefont
  {L\"ofwander}},\ }\bibfield  {title} {\enquote {\bibinfo {title} {Resonant
  second-harmonic generation in a ballistic graphene transistor with an
  ac-driven gate},}\ }\href {\doibase 10.1103/PhysRevB.93.035435} {\bibfield
  {journal} {\bibinfo  {journal} {Physical Review B}\ }\textbf {\bibinfo
  {volume} {93}},\ \bibinfo {pages} {035435} (\bibinfo {year}
  {2016}{\natexlab{a}})}\BibitemShut {NoStop}%
\bibitem [{\citenamefont {Korniyenko}\ \emph
  {et~al.}(2016{\natexlab{b}})\citenamefont {Korniyenko}, \citenamefont
  {Shevtsov},\ and\ \citenamefont {L\"ofwander}}]{Korniyenko16_2}%
  \BibitemOpen
  \bibfield  {author} {\bibinfo {author} {\bibfnamefont {Y.}~\bibnamefont
  {Korniyenko}}, \bibinfo {author} {\bibfnamefont {O.}~\bibnamefont
  {Shevtsov}}, \ and\ \bibinfo {author} {\bibfnamefont {T.}~\bibnamefont
  {L\"ofwander}},\ }\bibfield  {title} {\enquote {\bibinfo {title} {Nonlinear
  response of a ballistic graphene transistor with an ac-driven gate: High
  harmonic generation and terahertz detection},}\ }\href {\doibase
  10.1103/PhysRevB.94.125445} {\bibfield  {journal} {\bibinfo  {journal}
  {Physical Review B}\ }\textbf {\bibinfo {volume} {94}},\ \bibinfo {pages}
  {125445} (\bibinfo {year} {2016}{\natexlab{b}})}\BibitemShut {NoStop}%
\bibitem [{\citenamefont {Korniyenko}\ \emph {et~al.}(2017)\citenamefont
  {Korniyenko}, \citenamefont {Shevtsov},\ and\ \citenamefont
  {L{\"o}fwander}}]{Korniyenko17}%
  \BibitemOpen
  \bibfield  {author} {\bibinfo {author} {\bibfnamefont {Y.}~\bibnamefont
  {Korniyenko}}, \bibinfo {author} {\bibfnamefont {O.}~\bibnamefont
  {Shevtsov}}, \ and\ \bibinfo {author} {\bibfnamefont {T.}~\bibnamefont
  {L{\"o}fwander}},\ }\bibfield  {title} {\enquote {\bibinfo {title} {{Shot
  noise in a harmonically driven ballistic graphene transistor}},}\ }\href@noop
  {} {\bibfield  {journal} {\bibinfo  {journal} {Physical Review B}\ }\textbf
  {\bibinfo {volume} {95}},\ \bibinfo {pages} {165420--10} (\bibinfo {year}
  {2017})}\BibitemShut {NoStop}%
\bibitem [{\citenamefont {Pedersen}\ and\ \citenamefont
  {B{\"u}ttiker}(1998)}]{pedersen98}%
  \BibitemOpen
  \bibfield  {author} {\bibinfo {author} {\bibfnamefont {M.~H.}\ \bibnamefont
  {Pedersen}}\ and\ \bibinfo {author} {\bibfnamefont {M.}~\bibnamefont
  {B{\"u}ttiker}},\ }\bibfield  {title} {\enquote {\bibinfo {title} {Scattering
  theory of photon-assisted electron transport},}\ }\href@noop {} {\bibfield
  {journal} {\bibinfo  {journal} {Physical Review B}\ }\textbf {\bibinfo
  {volume} {58}},\ \bibinfo {pages} {12993} (\bibinfo {year}
  {1998})}\BibitemShut {NoStop}%
\bibitem [{\citenamefont {Platero}\ and\ \citenamefont
  {Aguado}(2004)}]{platero04}%
  \BibitemOpen
  \bibfield  {author} {\bibinfo {author} {\bibfnamefont {G.}~\bibnamefont
  {Platero}}\ and\ \bibinfo {author} {\bibfnamefont {R.}~\bibnamefont
  {Aguado}},\ }\bibfield  {title} {\enquote {\bibinfo {title} {Photon-assisted
  transport in semiconductor nanostructures},}\ }\href@noop {} {\bibfield
  {journal} {\bibinfo  {journal} {Physics Reports}\ }\textbf {\bibinfo {volume}
  {395}},\ \bibinfo {pages} {1--157} (\bibinfo {year} {2004})}\BibitemShut
  {NoStop}%
\bibitem [{\citenamefont {Kohler}\ \emph {et~al.}(2005)\citenamefont {Kohler},
  \citenamefont {Lehmann},\ and\ \citenamefont {H{\"a}nggi}}]{kohler05}%
  \BibitemOpen
  \bibfield  {author} {\bibinfo {author} {\bibfnamefont {S.}~\bibnamefont
  {Kohler}}, \bibinfo {author} {\bibfnamefont {J.}~\bibnamefont {Lehmann}}, \
  and\ \bibinfo {author} {\bibfnamefont {P.}~\bibnamefont {H{\"a}nggi}},\
  }\bibfield  {title} {\enquote {\bibinfo {title} {Driven quantum transport on
  the nanoscale},}\ }\href@noop {} {\bibfield  {journal} {\bibinfo  {journal}
  {Physics Reports}\ }\textbf {\bibinfo {volume} {406}},\ \bibinfo {pages}
  {379--443} (\bibinfo {year} {2005})}\BibitemShut {NoStop}%
\bibitem [{Note1()}]{Note1}%
  \BibitemOpen
  \bibinfo {note} {$T_n$ are computed from Eqs.~(A16)-(A20) in Ref.~\cite
  {Korniyenko16_2}, while $T_n'$ are computed from a set of equations that are
  obtained from Eqs.~(A16)-(A20) by first changing sign of $\protect
  \mathaccentV {vec}17EA$ and $\protect \mathaccentV {vec}17EC$, thereafter
  exchanging $\protect \mathaccentV {bar}016\eta \leftrightarrow \eta $, and
  finally exchanging superscripts $L\leftrightarrow R$.}\BibitemShut {Stop}%
\end{thebibliography}

%

\end{document}